\documentclass[review]{elsarticle}

\usepackage{hyperref}

\usepackage{amssymb}

\journal{Applied Mathematics Letters}

\usepackage{graphicx}
\usepackage{mathptmx}      
\usepackage{amssymb}
\usepackage[utf8]{inputenc}
\usepackage{bm}
\newcommand{\be}{\begin{equation}}
\newcommand{\ee}{\end{equation}}

\begin{document}

\begin{frontmatter}

\title{Asymptotically periodic points, bifurcations, and transition to chaos in fractional difference maps}

\author[mymainaddress,mysecondaryaddress]{Mark Edelman\corref{mycorrespondingauthor}}
\cortext[mycorrespondingauthor]{Corresponding author}
\ead{edelman@cims.nyu.edu}

\author[mymainaddress]{Avigayil Helman}  

\address[mymainaddress]{Stern College for Women, Yeshiva University, 245 Lexington Ave., New York, NY 10016}
\address[mysecondaryaddress]{Courant Institute of Mathematical Sciences at NYU, 251 Mercer Street, New York, NY 10012}

\begin{abstract}
In this paper we derive analytic expressions for coefficients of the equations that allow calculations of asymptotically periodic points in fractional difference maps. Numerical solution of these equations allows us to draw the bifurcation diagram for the fractional difference logistic map. Based on the numerically calculated bifurcation points, we make a conjecture that in fractional maps the value of the Feigenbaum constant $\delta$ is the same as in regular maps, $\delta=4.669...$.  
\end{abstract}


\end{frontmatter}


\section{Introduction}
\label{sec:1}

It is known that continuous and discrete fractional systems may
not have periodic solutions except fixed points (see, for example, \cite{PerD2,PerC2}). But the asymptotically periodic solutions do exist and the equations for finding these points in generalized fractional maps were derived in \cite{ME14,Helman}. These equations contain coefficients which are converging series. The numerical evaluation of these series, in the case fractional and fractional difference maps, can be reduced to the calculation of the Riemann $\zeta$-function. It is also known from the stability analysis of the discrete fractional systems (see \cite{MEstab}) that in the case of fractional difference maps the corresponding series are summable (see, e.g., \cite{AbuSaris2013,Cermak2015,Mozyrska2015,Anh}). In the following sections, after preliminaries in Section \ref{sec:2}, we derive the analytic expressions for the coefficients (sums) of the equations defining periodic points in the case of fractional difference maps in Section \ref{sec:3}. Then, in Section \ref{sec:4}, we present the bifurcation diagram for the order $\alpha=0.5$ fractional difference logistic map, which is based on the solutions of the equations defining the periodic points and compare this diagram with the bifurcation diagram obtained after 100000 iterations on a single trajectory. Based on the obtained numerically solutions for the bifurcation points, we draw a conjecture that the Feigenbaum number $\delta=4.669...$ has the same value for fractional difference maps.

\section{Preliminaries}
\label{sec:2}

For $0<\alpha<1$, the generalized universal $\alpha$-family of maps 
is defined as (see \cite{ME14,Helman}):
\begin{eqnarray}
x_{n}= x_0 
-\sum^{n-1}_{k=0} G^0(x_k) U_\alpha(n-k),
\label{FrUUMapN}
\end{eqnarray} 
where $G^0(x)=h^\alpha G_K(x)/\Gamma(\alpha)$, $x_0$ is the initial condition, $h$ is the time step of the map, $\alpha$ is the order of the map, $G_K(x)$ is a nonlinear function depending on the parameter $K$, $U_\alpha(n)=0$ for $n \le 0$, and $U_\alpha(n) \in \mathbb{D}^0(\mathbb{N}_1)$. The space $\mathbb{D}^i(\mathbb{N}_1)$ is defined as (see \cite{Helman})
{\setlength\arraycolsep{0.5pt}
\begin{eqnarray}
&&\mathbb{D}^i(\mathbb{N}_1)\ \ = \ \ \{f: |\sum^{\infty}_{k=1}\Delta^if(k)|>N, \ \ \forall N, \ \ N \in \mathbb{N}, 
\sum^{\infty}_{k=1}|\Delta^{i+1}f(k)|=C, \ \ C \in \mathbb{R}_+\},
\label{DefForm}
\end{eqnarray}
}
where $\Delta$ is a forward difference operator defined as
\begin{equation}
\Delta f(n)= f(n+1)-f(n).
\label{Delta}
\end{equation}
In the case Caputo fractional difference maps, which are defined as solutions of the Caputo h-difference equation \cite{DifSum,Fall,Chaos2014}
\begin{equation}
(_0\Delta^{\alpha}_{h,*} x)(t) = -G_K(x(t+(\alpha-1)h)),
\label{LemmaDif_n_h}
\end{equation}
where $t\in (h\mathbb{N})_{m}$, with the initial conditions 
 \begin{equation}
(_0\Delta^{k}_h x)(0) = c_k, \ \ \ k=0, 1, ..., m-1, \ \ \ 
m=\lceil \alpha \rceil,
\label{LemmaDifICn_h}
\end{equation}
the kernel $U_\alpha(n)$ is the falling factorial function: 
{\setlength\arraycolsep{0.5pt}
\begin{eqnarray}
&&U_{\alpha}(n)=(n+\alpha-2)^{(\alpha-1)} 
, \   \ U_{\alpha}(1)=(\alpha-1)^{(\alpha-1)}=\Gamma(\alpha).
\label{UnFrDif}
\end{eqnarray} 
}
The definition of the falling factorial $t^{(\alpha)}$ is
\begin{equation}
t^{(\alpha)} =\frac{\Gamma(t+1)}{\Gamma(t+1-\alpha)}, \ \ t\ne -1, -2, -3.
...
\label{FrFacN}
\end{equation}
The falling factorial is asymptotically a power function:
\begin{equation}
\lim_{t \rightarrow
  \infty}\frac{\Gamma(t+1)}{\Gamma(t+1-\alpha)t^{\alpha}}=1,  
\ \ \ \alpha \in  \mathbb{R}.
\label{GammaLimitN}
\end{equation}
The $h$-falling factorial $t^{(\alpha)}_h$ is defined as
\begin{eqnarray}
&&t^{(\alpha)}_h =h^{\alpha}\frac{\Gamma(\frac{t}{h}+1)}{\Gamma(\frac{t}{h}+1-\alpha)}= h^{\alpha}\Bigl(\frac{t}{h}\Bigr)^{(\alpha)}, \  \
\frac{t}{h} \ne -1, -2, -3,
....
\label{hFrFacN}
\end{eqnarray}
 
The following equations define period-$l$ points in generalized fractional maps \cite{ME14}
{\setlength\arraycolsep{0.5pt}   
\begin{eqnarray} 
&&x_{lim,m+1}-x_{lim,m}=S_1 G^0(x_{lim,m})+\sum^{m-1}_{j=1}S_{j+1} G^0(x_{lim,m-j})
\nonumber \\
&&+\sum^{l-1}_{j=m}S_{j+1} G^0(x_{lim,m-j+l}), \  \ 0<m<l,
\label{LimDifferences}
\\
&&\sum^{l}_{j=1} G^0(x_{lim,j})=0,
\label{LimDifferencesN}
\end{eqnarray}
}
where
{\setlength\arraycolsep{0.5pt}   
\begin{eqnarray} 
&&S_{j+1}=\sum^{\infty}_{k=0}\Bigl[
U_{\alpha} (lk+j) - U_{\alpha} (lk+j+1)\Bigr], \  \ 0 \le j<l.
\label{Ser}
\end{eqnarray}
}
It is easy to see that 
\begin{equation}
\sum^{l}_{j=1}S_j=0.
\label{Ssum}
\end{equation}

\section{Sums $S_p$ for $l$-cycles of fractional difference maps}
\label{sec:3}

We will call $S_p$ sum for an $l$-cycle $S_{p,l}$.
The definition of $S_{p,l}$ from \cite{ME14} for fractional difference maps may be rewritten using the following chain of transformations:
{\setlength\arraycolsep{0.5pt}   
\begin{eqnarray} 
&&S_{p,l}= 
\sum^{\infty}_{k=0}\Bigl[(lk+p+\alpha-3)^{(\alpha-1)}- (lk+p+\alpha-2)^{(\alpha-1)} \Bigr] =\sum^{\infty}_{k=0}
\Bigl[\frac{\Gamma(lk+p+\alpha-2)}{ \Gamma(lk+p-1)}
\nonumber \\
&&- \frac{\Gamma(lk+p+\alpha-1)}
{ \Gamma(lk+p)} \Bigr] =(1-\alpha)\sum^{\infty}_{k=0}
\frac{\Gamma(lk+p+\alpha-2)}{\Gamma(lk+p)}
=-\Gamma(\alpha) \sum^{\infty}_{k=0}
\frac{\Gamma(lk+p+\alpha-2)}{\Gamma(\alpha-1)\Gamma(lk+p)}
\nonumber \\
&&=-\Gamma(\alpha) \sum^{\infty}_{k=0}
\left( \begin{array}{c}
lk+p+\alpha-3 \\ lk+p-1
\end{array} \right)
=\Gamma(\alpha) (-1)^{p}\sum^{\infty}_{k=0}(-1)^{lk}
\left( \begin{array}{c}
1-\alpha \\ lk+p-1
\end{array} \right).
\label{Sp_l}
\end{eqnarray}
}
Using absolute convergence of series and the following identity (see \cite{MSS})
{\setlength\arraycolsep{0.5pt}   
\begin{eqnarray} 
&&\sum^{\infty}_{k=0}
\left( 
\begin{array}{c}
\gamma \\ t+ks
\end{array} \right)
= \frac{1}{s}\sum^{s-1}_{j=0} \omega^{-jt}(1+\omega^j)^\gamma,
\label{I1_3}
\end{eqnarray}
}
where $\omega=e^{i2\pi/s}$, for the even and odd periods we obtain
{\setlength\arraycolsep{0.5pt}   
\begin{eqnarray} 
&&S_{p,2n}= \Gamma(\alpha) (-1)^{p}\sum^{\infty}_{j=0}
\left( \begin{array}{c}
1-\alpha \\ 2nj+p-1
\end{array} \right)
=\frac{\Gamma(\alpha)}{2n}(-1)^p\sum^{2n-1}_{j=0}e^{-i\pi j(p-1)/n}(1+ e^{i\pi j/n})^{1-\alpha}
\nonumber \\
&&=\frac{\Gamma(\alpha)}{2n}(-1)^p\Bigl[2^{1-\alpha}
+\sum^{n-1}_{j=1}\Bigl(e^{-i\pi j(p-1)/n} e^{i\pi j(1-\alpha)/(2n)}(2\cos(\pi j/(2n)))^{1-\alpha}
\nonumber \\
&&+ e^{i\pi j(p-1)/n} e^{-i\pi j(1-\alpha)/(2n)} (2\cos(\pi j/(2n)))^{1-\alpha}\Bigr) \Bigr]
\nonumber \\
&&=\frac{\Gamma(\alpha) 2^{-\alpha}}{n}(-1)^p\Bigl[1
+2\sum^{n-1}_{j=1}(\cos(\pi j/(2n)))^{1-\alpha}\cos(\pi j(2p+\alpha-3)/(2n)) \Bigr],
\label{Sp_2n}
\end{eqnarray}
}
{\setlength\arraycolsep{0.5pt}   
\begin{eqnarray} 
&&S_{p,2n+1}= \Gamma(\alpha) (-1)^{p}\sum^{\infty}_{j=0}(-1)^j
\left( \begin{array}{c}
1-\alpha \\ (2n+1)j+p-1
\end{array} \right)
\nonumber \\
&&=\Gamma(\alpha) (-1)^{p} \Bigl\{ \sum^{\infty}_{j=0}
\left( \begin{array}{c}
1-\alpha \\ 2(2n+1)j+p-1
\end{array} \right)
-\sum^{\infty}_{j=0}
\left( \begin{array}{c}
1-\alpha \\ 2(2n+1)j+p+2n
\end{array} \right)
\Bigr\}
\nonumber \\
&&=\frac{\Gamma(\alpha)}{2(2n+1)}(-1)^p \Bigl\{   \sum^{4n+1}_{j=0}e^{-\frac{i\pi j(p-1)}{2n+1}}(1+ e^{\frac{i\pi j}{2n+1}})^{1-\alpha}-\sum^{4n+1}_{j=0}e^{-\frac{i\pi j(p+2n)}{2n+1}}(1
\nonumber \\
&&+ e^{\frac{i\pi j}{2n+1}})^{1-\alpha}\Bigr\}
=\frac{\Gamma(\alpha)}{2(2n+1)}(-1)^p    \sum^{2n}_{j=1} \Bigl\{      \Bigl(e^{-\frac{i\pi j(p-1)}{2n+1}}- e^{-\frac{i\pi j(p+2n)}{2n+1}}\Bigr)     (1
\nonumber \\
&&+ e^{\frac{i\pi j}{2n+1}})^{1-\alpha}
+\Bigl(e^{\frac{i\pi j(p-1)}{2n+1}}- e^{\frac{i\pi j(p+2n)}{2n+1}}\Bigr)     (1+ e^{-\frac{i\pi j}{2n+1}})^{1-\alpha}\Bigr\}
\nonumber \\
&&=\frac{\Gamma(\alpha)}{2n+1}(-1)^p i    \sum^{2n}_{j=1}\Bigl(2\cos\frac{\pi j}{2(2n+1)} \Bigr)^{1-\alpha}
\sin\frac{j\pi}{2}\times
\Bigl(e^{-\frac{i\pi j(2p+2n-2+\alpha)}{2(2n+1)}}
\nonumber \\
&&- e^{\frac{i\pi j(2p+2n-2+\alpha)}{2(2n+1)}}\Bigr)
=\frac{2^{2-\alpha}\Gamma(\alpha)}{2n+1}(-1)^p   
\sum^{2n}_{j=1}\Bigl(\cos\frac{\pi j}{2(2n+1)} \Bigr)^{1-\alpha}
\sin\frac{j\pi}{2}
\nonumber \\
&&\times
\sin\frac{\pi j(2p+2n-2+\alpha)}{2(2n+1)}
\nonumber \\
&&=\frac{2^{2-\alpha}\Gamma(\alpha)}{2n+1}(-1)^p   
\sum^{n-1}_{j=0}\Bigl(\cos\frac{\pi (2j+1)}{2(2n+1)} \Bigr)^{1-\alpha}
(-1)^j
\nonumber \\
&&\times
\sin\frac{\pi (2j+1)(2p+2n-2+\alpha)}{2(2n+1)}
\label{Sp_2n+1}
\end{eqnarray}
}
In \cite{MSS}, Eq.~(\ref{I1_3}) is proven for integer values of $\gamma \in \mathbb{N}_0$. The question is whether this identity can be extended to $\gamma \in \mathbb{R}$. The answer is NO. The calculated values of $S_{2,4}$ obtained when we used Eq.~(\ref{Sp_2n}) for $\alpha=0.5$, $\alpha=0.99$, and $\alpha=0.999$ are 1.309697, 0.257228, and 0.2507115. The corresponding values obtained using the exact expression (see Eq.~(35) in \cite{ME14}) are 1.029970, 0.2571808, and 0.2507111.

\section{Bifurcation diagrams for the fractional difference logistic map with $0<\alpha<1$}
\label{sec:4}

\begin{figure}[!t]
\includegraphics[width=1.0 \textwidth]{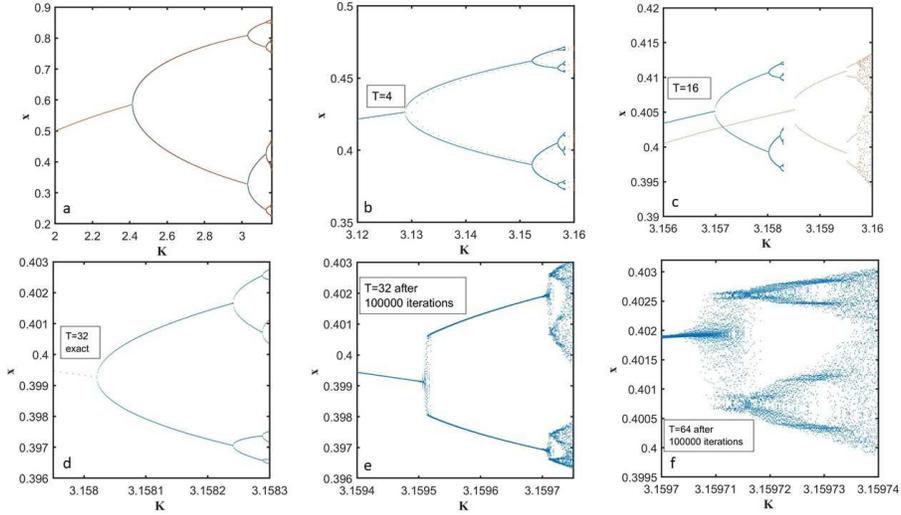}
\vspace{-0.25cm}
\caption{A part of the bifurcation diagram for the Caputo fractional difference logistic map of the order $\alpha=0.5$ from $K=2$ (fixed point) to approximately $K=3.16$ (T=128 periodic point). In figures a-c the steady line represents the solution of  Eqs.~(\ref{LimDifferences})~and~(\ref{LimDifferencesN}) (the exact solution) and the dots represent numerical calculations on a single trajectory with $x_0=0.3$ after $10^5$ iterations. 
Figure d represents the exact solution from $T=32$ on the left to $T=128$ on the right. Figures e and f represent the calculations on a single trajectory.
}
\label{fig1}
\end{figure}

In the generalized fractional logistic map 
\begin{eqnarray}
G_K(x)=x-Kx(1-x).
\label{LM}
\end{eqnarray} 
We solved Eqs.~(\ref{LimDifferences})~and~(\ref{LimDifferencesN}) numerically using Mathematica's Newton's Method (the exact solution) to draw bifurcation diagrams for various values of $\alpha \in (0,1)$ (we assume $h=1$). Our results and Mathematica codes are available upon request. Here, as an example, we present only the results for the value of $\alpha=0.5$, which is far from the integer values zero and one.
Fig.~\ref{fig1} shows the results of our calculations for $\alpha=0.5$.
From Fig.~\ref{fig1}~b one may see that already for the period four ($T=4$) points there is a noticeable difference between the exact solutions and the results obtained after $10^5$ iterations on a single trajectory. For $T>16$ (Figs.~\ref{fig1}~b,~e,~f) the results obtained after $10^5$ iterations on a single trajectory seem to be noticeably inaccurate. Multiple papers investigating various fractional difference maps contain bifurcation diagrams obtained by iterations on a single trajectory and the number of iterations in all these papers is much less than $10^5$.

Table~\ref{table:T3_Fr} shows the bifurcation points $K(n)$ (the period $2^{n-1}$ -- period $2^{n}$ bifurcations) obtained using Eqs.~(\ref{LimDifferences})~and~(\ref{LimDifferencesN}) and the corresponding values of the ratios $\delta_1=\Delta K(n-2)/\Delta K(n-1)= [K(n-1)-K(n-2)]/[K(n)-K(n-1)]$ for the order $\alpha=0.5$ fractional difference logistic map. The corresponding values for the regular logistic map are taken from Wikipedia \cite{FC}.
\begin{table}[ht!]
\centering
    \begin{tabular}{| c  | c   |  c  |  c    |      c  | c      | c       |}
    \hline 
    $n (T=2^n)$ & $K_1(n)$ & $K_{.5}(n)$ & $\frac{\Delta K_{1}(n-2)}{\Delta K_{1}(n-1)}$  &  $\frac{\Delta K_{.5}(n-2)}{\Delta K_{.5}(n-1)}$ & $\delta_1-\delta_F$ & $\delta_{.5}-\delta_F$  \\ \hline
    1          & 3         & 2.4142136 &          &          &         & \\ \hline
    2          & 3.4494897 & 3.0315081 &          &          &         & \\ \hline
    3          & 3.5440903 & 3.1288030 & 4.751447 & 6.345 & .082245 & 1.6754 \\ \hline
    4          & 3.5644073 & 3.1522546 & 4.656229 & 4.149 &-.012973 &
-.5204  \\ \hline
    5          & 3.5687594 & 3.1569819 & 4.668321 & 4.961 &-.000881 &     .2917  \\ \hline
    6          & 3.5696916 & 3.1580209 & 4.668633 & 4.550 &-.000569 & -.1193  \\ \hline
    7          & 3.5698913 & 3.1582410 & 4.668002 & 4.721 &-.001200 & .0514  \\ \hline
    8          & 3.5699340 & 3.1582884 & 4.676815 & 4.6435 & .007613 & -.0257 \\ \hline
    9          &           & 3.15829845 &          & 4.716 &         & .0472  \\ \hline
    10         &           & 3.15830062 &          & 4.631 &         & -.0379 \\ \hline
    \end{tabular}
    \caption{Approaching the Feigenbaum constant $\delta$ in the regular and fractional difference logistic maps. $K_1$ is the period $2^{n-1}$ -- period $2^{n}$ bifurcation point in the regular logistic map ($\alpha=1$); $K_{.5}$ is the same point in the case $\alpha=.5$; $\delta_1=\Delta K_{1}(n-2)/\Delta K_{1}(n-1)= [K_{1}(n-1)-K_1(n-2)]/[K_{1}(n)-K_1(n-1)]$; $\delta_{.5}$ is the same value in the case $\alpha=.5$; $\delta_F=\delta=4.6692016$.}
    \label{table:T3_Fr}
\end{table}
As in the case of the regular logistic map, the values of $\delta$ oscillate around the Feigenbaum number but converge significantly slower. The slow convergence is expected because, in general, the convergence in fractional maps follows the power law while the convergence in regular maps is exponential. From the authors' point of view, the data present sufficient evidence to make a conjecture that the Feigenbaum number $\delta$ exists in fractional difference maps and has the same value. 

\section{Conclusion}
\label{sec:5}
In this paper we derived the analytic expressions for the coefficients of the equations that define periodic points in fractional difference maps of the orders $0<\alpha<1$ (Eqs.~(\ref{Sp_2n})~and~(\ref{Sp_2n+1})). We draw the bifurcation diagram for the fractional difference logistic map of the order $\alpha=0.5$ which is based on the solution of Eqs.~(\ref{LimDifferences})~and~(\ref{LimDifferencesN}) and  
showed that bifurcation diagrams based on the iterations on a single trajectory contain significant errors. 
The data in Table~\ref{table:T3_Fr} present sufficient evidence to make a conjecture that the Feigenbaum number $\delta$ exists in fractional difference maps and has the same value as in regular maps. We believe that it is worthwhile to invest time and to apply efforts to prove this conjecture.

\section*{Data availability}

Data will be made available on request.

\section*{Acknowledgements}
The first author acknowledges support from Yeshiva University's 2021--2022 Faculty Research Fund, expresses his gratitude to the administration of Courant Institute of Mathematical Sciences at NYU
for the opportunity to perform some of the computations at Courant,
and expresses his gratitude to Virginia Donnelly for technical help.


\end{document}